\documentclass[sigconf]{acmart}
\AtBeginDocument{%
  \providecommand\BibTeX{{%
    \normalfont B\kern-0.5em{\scshape i\kern-0.25em b}\kern-0.8em\TeX}}}

\setcopyright{none}
\usepackage{_macros}
\usepackage{_custom}

\begin{document}

\hidecomments
% \showrevisions{REVBLUE}

\raggedbottom

\title{
{Language Scent: Exploring Cross-Language Information~Navigation}
}

% \affiliation{%

\author{Jiawen Stefanie Zhu}
\orcid{0009-0002-2652-7241}
\affiliation{
  \institution{University of Washington}
  \city{Seattle}
  \state{Washington}
  \country{USA}
}
\email{jiawenz2@uw.edu}

\author{Katharina Reinecke}
\orcid{0000-0001-7897-9325}
\affiliation{
  \institution{University of Washington}
  \city{Seattle}
  \state{Washington}
  \country{USA}
}
\email{reinecke@cs.washington.edu}

\author{Tanushree Mitra}
\orcid{0000-0002-9507-6192}
\affiliation{
  \institution{University of Washington}
  \city{Seattle}
  \state{Washington}
  \country{USA}
}
\email{tmitra@uw.edu}

% Use these to make cleaner submission without weird fake names
% \acmConference[]{}{}{}
% \acmYear{}
% \copyrightyear{}
% \acmPrice{}
% \acmDOI{}
% \acmISBN{}
% \setcopyright{none}

% actual info
\copyrightyear{2026}
\acmYear{2026}
\setcopyright{cc}
\setcctype{by-nc-nd}
\acmConference[UIST '26]{The 39th Annual ACM Symposium on User Interface Software and Technology}{November 02--05, 2026}{Detroit, MI, USA}
\acmBooktitle{The 39th Annual ACM Symposium on User Interface Software and Technology (UIST '26), November 02--05, 2026, Detroit, MI, USA}
\acmDOI{10.1145/3830398.3830488}
\acmISBN{979-8-4007-2856-3/2026/11}


\begin{abstract}

%% v4
While multilingual users often switch between languages when seeking information, {this process remains undersupported by current systems where information is typically siloed by language.}
Our formative study reveals that users select their search language based on its perceived value for their current information need, a concept we formalize as \textbf{language scent}.
\rev{Language scent extends Pirolli and Card's information foraging theory --- which explains how users navigate among already-encountered sources --- to the multilingual case, where users' choice of query language determines which sources they can encounter in the first place.
Building on this insight, we designed \sys{}, a multilingual information seeking system that provides proximal cues for gauging the language scent of different languages.}
Finally, we conducted a lab study with 16 multilingual speakers to understand \sys{}'s utility, usage patterns and application contexts.
\end{abstract}




\begin{teaserfigure}
\centering
    \includegraphics[width=1\linewidth]{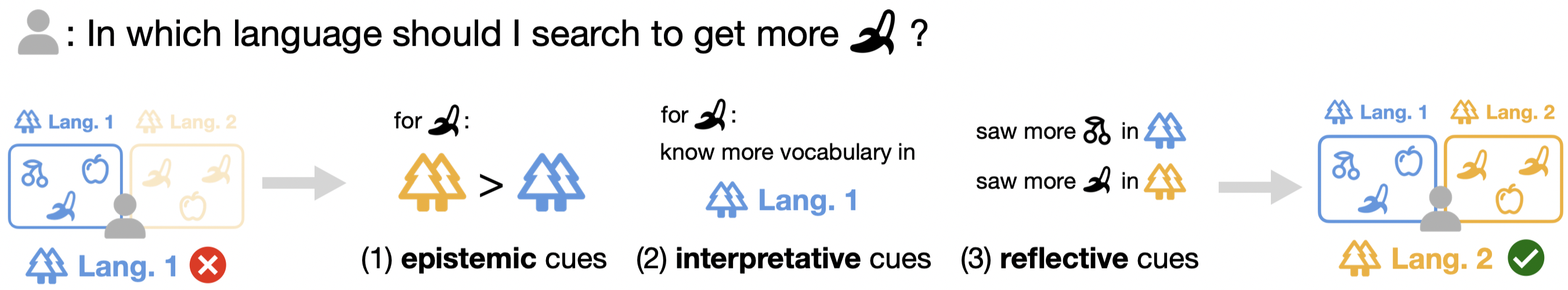}
    \caption{
    % Overview of language scent and \sys{}.
    \rev{
    In multilingual settings, information is siloed by language ("forests"), and the search language chosen determines which sources ("fruits") are accessible. Users typically only see results in their search language, with little awareness of what exists in other language subspaces. 
    \sys{} helps users navigate across these subspaces using proximal cues: epistemic cues convey the information value of subspaces, interpretative cues convey users' ability to extract that information (e.g., language proficiency), and reflective cues support learning from past choices.
    }
    }
    \label{fig:teaser}
\end{teaserfigure}

\maketitle

\section{Introduction}
\label{sec:intro}
Information sources across languages often contain distinct content and framing.
Consulting multiple languages when information seeking can thus surface complementary perspectives and lead to a more comprehensive understanding \cite{quelleLostTranslationMultilingual2023, hecht2010tower, bao2012omnipedia}.
Multilingual speakers, for one, often draw on multiple languages to meet their information needs \cite{maftoon2011bilingual, gaoTakingLanguageDetour2022, aula2009MultilingualSearchStrategiesb, steichen2021HowMultilingualUsersa}.
However, current search systems tend to silo information by language, for example by showing only sources in the query language, making this broader information space difficult to navigate.
Overall, cross-lingual information seeking remains under-supported.

While a large body of prior work has examined multilingual speakers' information seeking behaviours --- including their language preferences across topics \cite{aula2009MultilingualSearchStrategiesb, gaoTakingLanguageDetour2022, wang2018SwitchingLanguagesOnline}, how they navigate different multilingual search result interfaces \cite{steichen2021HowMultilingualUsersa, steichenSupportingModernPolyglot2015}, and how they (re)formulate queries when switching language \cite{fu2018MixedLanguageQueries, fuQueryReformulationPatterns2017} --- these studies tend to operate under the implicit assumption that users \textit{inherently know} which language is best to use for a given search task.
However, closer examination suggests that this assumption may not always hold.
Empirical findings frequently highlight trial-and-error processes, wherein users recognize the need to switch to another language only after extensive, unproductive searches in their initial language \cite{aula2009MultilingualSearchStrategiesb, zhang2024HowBridgeGap}.
Despite this, the mechanisms by which users determine appropriate search languages and the challenges they face in doing so remain poorly understood.

To bridge this gap, we investigate the following research questions:
\begin{enumerate}[label=\textbf{RQ\arabic*}, topsep=5pt]
    \item How do multilingual users strategize and navigate between languages during information seeking?
    \item How do we design search systems that facilitate this cross-language strategy formation and refinement process?
    \item How do such designs affect users’ information-seeking processes and outcomes?
\end{enumerate}
We focused on English-Chinese users as a case study, following Human-Computer Interaction (HCI) conventions of studying a specific language pair to be able to capture nuances \cite{bawaMultilingualUsersPrefer2020, choiMultilingualConversationalAgent2023, xiao2025SustainingHumanAgencyc, aula2009MultilingualSearchStrategiesb}.

We began by conducting a formative study (\textbf{RQ1}) with $10$ English-Chinese speakers to understand their information seeking strategies and pain points, especially with regard to cross-language information navigation.
\rev{We found that participants selected their search language based on the {perceived value of each language} for fulfilling their current information need, which was influenced by factors such as the perceived informational value of sources in each language, topic-specific language proficiency, and the operational costs of switching languages within a system.
We see parallels between this decision mechanism and information foraging theory (IFT), which theorizes that users navigate between sources guided by information scent --- the {perceived value of each source} --- shaped by ``imperfect information at intermediate locations'' like thumbnails or links \cite{pirolli1999InformationForaginga}.
While information scent explains how users navigate among \textit{encountered} sources, IFT does not explicitly consider which sources enter users' consideration in the first place.
This gap is critical in multilingual settings, where sources are siloed by language, and the choice of search language determines which sources users can encounter at all.
Hence, we introduce the concept of \textbf{language scent} as the \textbf{perceived value of searching in a particular language}, to extend IFT to contexts where the information space is fragmented across multiple siloed subspaces.
Language scent determines which language subspace a user chooses to search; information scent then determines how users navigate within that chosen subspace.
}

\rev{Language scent provides a conceptual lens for explaining multilingual information seeking behaviours.
For example, the reason users in current systems often continue searching in a single language longer than is productive is that the lack of proximal cues about other language subspaces forces them to rely solely on imperfect prior experiences and intuitions, preventing them from developing accurate language scent that effectively leads them to relevant language subspaces.
}
To address this gap, we derived design guidelines from formative study results to develop \sys{} (\textbf{RQ2}), \rev{a multilingual information seeking system designed to enable language-scent-guided information navigation.}
\rev{\sys{} provides various types of proximal cues for gauging the language scent of different languages.
Epistemic cues like comparative summaries and previews of information in other languages act as intermediate representations of the relevant information in each language subspace.
Interpretative cues like multilingual displays and cross-lingual keywords, in turn, connect distal information to users' existing knowledge, helping them assess how efficiently they can extract information in each language.
Finally, \sys{} provides reflective cues on past search activity with respect to language, helping users understand and refine their strategies and judgment of language scent. 
}

We examined how \sys{} influences users' cross-language information seeking experience in a lab study with 16 English-Chinese speakers \rev{(\textbf{RQ3})}.
Results showed that \sys{} helped users develop more flexible and granular search strategies, enabling them to gather a more diverse set of information.
Overall, our findings suggest that language scent is a valuable lens on cross-language information navigation, and offer insights for the research and design of future multilingual search systems.

\section{Related Work}

\subsection{{Grounding in Information Foraging Theory}}
Information foraging theory (IFT) is a fundamental theory of information seeking, positing that people navigate information spaces by using ``imperfect information at intermediate locations'' --- such as thumbnails or links --- to gauge the value of each source and choose accordingly \cite{pirolli1999InformationForaginga}.
This perception of the value of a source based on proximal cues is called its information scent \cite{pirolli1999InformationForaginga}.
Information scent, in its original formulation \cite{pirolli1999InformationForaginga}, assumes a monolingual context and focuses on navigating between \rev{\textit{encountered}} patches of information within a coherent information space.
However, IFT does not account for the multilingual context, where \rev{users must navigate multiple siloed subspaces fragmented by language}.
This introduces an additional decision layer in the information-seeking process: \rev{Users must first choose the appropriate subspace, which determines the sources that could be encountered, before navigating between sources.}
\rev{
Focusing on scent as the perception of proximal cues signalling distal value, we defined language scent as the perceived value of searching in a particular language to study this extra layer of decision-making.
Language scent extends information scent theory by accounting for an information topology that is fragmented by language rather than a coherent whole. 
This lets us study the decision mechanism by which users select a language subspace (guided by language scent), which precedes source navigation (guided by information scent).
}

Existing systems work has explored ways of enhancing users’ (general) information scent.
One line of work focuses on {designing proximal cues that amplify the signal of individual sources}, for example through enhanced thumbnail images \cite{woodruff2001UsingThumbnailsSearch, taiebmaimon2025EnhancingSearchVisualizations}.
A recent line of work surfaces more distant information patches by suggesting search queries {as proximal cues} based on users’ past interactions \cite{palani2021CoNotateSuggestingQueries, palani2022InterWeavePresentingSearcha}.
However, these were designed with monolingual contexts in mind and did not consider the additional layer of navigating between multiple information subspaces introduced by multilingual information seeking.
\sys{} addresses this gap by designing for language scent to support the meta-level decision-making process of switching between languages.

\subsection{{Understanding Cross-lingual Information Seeking}}

Information seeking is the conscious effort of acquiring information to fill a need or gap in one's knowledge \cite{case2016LookingInformationSurvey}, and is an important activity in daily life \cite{athukorala2016ExploratorySearchDifferent}.

% gap 1: decision-making process
Existing work has investigated how multilingual users in particular seek information, and showed that they leverage different languages during this process {\cite{aula2009MultilingualSearchStrategiesb, gaoTakingLanguageDetour2022, wang2018UnderstandingMultilingualWeb}}.
One line of work investigates patterns used by people when employing certain languages. For example, users switch between their country-of-residence language and their native language during crisis information seeking to balance digestibility and authenticity \cite{gaoTakingLanguageDetour2022}.
Another line of work focuses on specific challenges and strategies that people encounter when switching languages, for example query (re)formulation \cite{fuQueryReformulationPatterns2017, fu2018MixedLanguageQueries, albarillo2018InformationCodeSwitchingStudy}.
However, none of the existing work has examined the process \textit{before} users decide to use a certain language or switch languages, i.e. their search strategy formation stage.
Rather, there is an implicit assumption that multilingual users always know when to use which language.
This is not necessarily true, since existing work suggests that multilingual users often rely on trial and error, sticking with one language until they realize it is unable to satisfy their information needs \cite{aula2009MultilingualSearchStrategiesb}.
{Our formative study fills this gap by examining how multilingual users form and refine their information-seeking strategies, through the new lens of language scent.}

\subsection{Supporting Cross-lingual Information Seeking}
While there is an abundance of work on underlying cross-lingual retrieval algorithms \cite{hull1996QueryingLanguagesDictionarybaseda, peters2012MultilingualInformationRetrieval}, there is limited work on {user-facing} multilingual information seeking {tools and systems}.
One such line of work primarily focuses on supporting query reformulation \cite{fuQueryReformulationPatterns2017, steichen2021HowMultilingualUsersa}, {for example by automatically adapting imperfect user queries into more effective versions \cite{sun2023CLQRCrossLingualEnhanced}.}
{Another line examines how to design search result page UI to organize results from multiple languages effectively \cite{chuTranSearchMultilingualSearch2017, lingComparativeUserStudy2018, steichenMultilingualNewsSearch2023}.
Neither supports the strategic navigation across languages.
Our system, \sys{}, \rev{fills this gap by supporting language-scent-guided information navigation, helping users better explore information from multiple languages.}
}

%While they are not tailored for multilingual users, 
{There are general information seeking tools that are related to some of our high-level design goals, like reducing information overload and facilitating search strategy formation.}
For example, systems like DiscipLink \cite{zhengDiscipLinkUnfoldingInterdisciplinary2024a} and Selenite \cite{liu2024SeleniteScaffoldingOnline} consolidate and organize raw information, while CoNotate \cite{palani2021CoNotateSuggestingQueries} and InterWeave \cite{palani2022InterWeavePresentingSearcha} suggest relevant queries given user contexts.
{However, our contribution does not lie in general-purpose consolidation or suggestion mechanisms, but in manifesting the design concept of surfacing language scent through \sys{} to better assist multilingual users.}

\subsection{Characterizing Multilingual Users}
\label{sec:musers}
Multilingualism is an overloaded term with many definitions and interpretations \cite{coulmas2018IntroductionMultilingualismLanguage, mackey1962DescriptionBilingualism, maftoon2011bilingual}.
Even in HCI alone, multilingualism carries two distinct connotations, one emphasizing the multi-competence of knowing multiple languages, e.g. \cite{bawaMultilingualUsersPrefer2020}, and the other emphasizing not being a native speaker of English, e.g. \cite{kimItsMyLanguage2024}.
In this work, we define multilingual users as people who are fluent, i.e. can produce ``complete meaningful utterances'' \cite{maftoon2011bilingual}, in two or more languages, focusing on the multi-competence aspect.

Furthermore, studying multilingual users as a whole is rare, given the diversity within this population.
Rather, HCI research conventionally focuses on specific language pairs as case studies, enabling a more nuanced understanding \cite{bawaMultilingualUsersPrefer2020, choiMultilingualConversationalAgent2023, xiao2025SustainingHumanAgencyc, aula2009MultilingualSearchStrategiesb}. 
We focus on English and Chinese in this project, since they rank as the top two most spoken languages globally \cite{topLang} and are spoken by some members of the research team \cite{gaoTakingLanguageDetour2022}.
\section{Formative Study}

We conducted a formative study to investigate how multilingual users navigate between languages during information seeking.

\subsection{Method}

The study was conducted as an online interview study with 10 participants (\textbf{P\#}, 7 women, 3 men; mean age $= 25 \pm 3$), recruited through social media and snowball sampling.
All participants were native speakers of Chinese and at least independent users of English according to the Common European Framework of Reference for Languages (CEFR) \cite{CEFRLevelsCommona}.
Specifically, one participant was at the B-level (independent users), nine were at the C-level (proficient/near-native users).
\rev{All participants lived in a Chinese-speaking region until at least age 15 ($M = 18.6, SD = 2.5$; 1 in Taiwan and Hong Kong, 9 in Mainland China) and, at the time of the study, resided in an English-speaking country (5 in Canada, 5 in the United States).}
All reported regularly seeking information online, on average about once a day (mean $= 0.97 \pm 0.09$ times).
\rev{Additionally, on average participants rated their frequency of information seeking in Chinese as $5.77$ and in English as $7.00$ on a 7-point Likert scale ($1 = \text{Very Rarely}$, $7 = \text{Very Frequently}$).}

As a warm-up exercise, and to observe users’ multilingual information seeking behaviours, we selected two tasks likely to induce language switching, based on prior work \cite{steichenMultilingualNewsSearch2023}: exploring public opinions on (1) the release of the AI model DeepSeek and (2) the practice of vegetarianism.
Participants had 10 minutes for each task and were asked to think aloud \cite{ThinkingAloud1, charters2003UseThinkaloudMethods} during the process.
We concluded with a 30--40 minute semi-structured interview.
\rev{The interview focused on participants' (1) language choice and switching strategies during information seeking, (2) perceptions of current tools' support for their workflow, and (3) desired features and envisioned designs.}
In total, each study session took approximately 60 minutes and participants were remunerated US \$15.
This study was approved by our institution's ethics review board.

Task sessions and interviews were screen- and audio-recorded.
We conducted an inductive thematic analysis by open-coding the interview transcripts \cite{braunUsingThematicAnalysis2006a}, \rev{focusing on language choice in multilingual search.
The first author identified initial themes, which were refined with co-authors until consensus.}
\rev{Screen and audio recording were used when context was needed (e.g. resolve ambiguity).}

\subsection{Findings}
\label{sec:scent}

\rev{Our formative study provided the empirical grounding for language scent (Section \ref{sec:intro}):
multilingual users navigate siloed language subspaces based on the perceived value of each language for their current information seeking goal.
}

\rev{We identified factors shaping language scent, extracted the challenges arising from a lack of proximal language-related cues in existing systems, and derived design guidelines for systems that enable language-scent-guided search.
}

\subsubsection{Understanding Language Scent}

\rev{Participants chose the language for information seeking based on its perceived value.
In current systems, this value judgment was primarily shaped by intuition and prior information-seeking experience, since systems provided no proximal cues about language subspaces.}
For example, while some participants may use English to \pqt{[learn] a concept for school}{P8}, or due to the availability of a \pqt{wide variety of sources}{P3}, they might instead decide to search in \pqt{Chinese for creating travel plans for more practical tips}{P9}.
\rev{This example illustrates some of the factors shaping language scent.
We systematically unpack these factors} --- \epistemic{}, \interpretative{}, and \practical{} --- in the subsections below, along with the challenges (\textbf{C\#}) users face when current systems fail to support accurate language scent.

\textbf{\Epistemic{} Factors.}
\fixme{We identified aspects of language scent related to the informational content available in a given language, including its availability, quality, and framing.} 
Participants viewed language as a proxy for \pqt{see[ing] other perspectives}{P1}, explaining that \pqt{even if [they] ask the same question}{P10}, \pqt{most likely the things [they] get from searching in Chinese and the things [they] get from searching in English are different}{P10}.
Users also considered the applicability of information to their own context and positionality.
For example, P4 explains that to \pqt{know about the general visa application process}{P4}, it doesn't matter which language they use, but only \pqt{the Chinese side is going to tell you to make your resume a little less sensitive to make sure it doesn't raise any eyebrows}{P4}.

Participants began with a prior \rev{expectation} of the value of searching in each language to guide their search actions, refining this model \pqt{if the acquired information is different from what [they] thought before}{P1}.
However, based on their prior expectations alone, participants \pqt{often [did not know] which language is better at first}{P3}, and even if they did, found it difficult to predict \pqt{if [they]'re getting the kind of information [they] want by searching in [a particular] language}{P3}.
\fixme{This difficulty is compounded by current systems}, which silo information by language and provide few cues to help users develop more accurate intuitions about what each language contains (\textbf{C1 — Limited Awareness of Information in \rev{Each Language}}).
A related phenomenon is that users often \rev{do not pay attention to language-related aspects of their information seeking behaviours and strategies. (\textbf{C2 — {Limited Awareness of Their Own Cross-Language Navigation Patterns}})
This leads to unproductive trial-and-error processes, where users only start questioning their approach and adapting their strategies after many unfruitful searches.
}

\textbf{\fixme{\Interpretative{}} Factors.}
\fixme{These are aspects of language scent related to the (perceived) ease of processing or digesting information using a particular language, shaped by the user’s current language environment, past experiences, and language skills.
} 
Participants explained that the language they choose during information seeking is also \pqt{about [their] thinking process}{P9} and frequently \pqt{not really a rational categorization}{P2}.
Users would often \pqt{just use the language that comes to mind first}{P9}, \pqt{recall where [they] first encountered this problem and then habitually rely on that path}{P2}, or default to the language of greatest proficiency, rather than \fixme{deliberately} considering which language would best satisfy their information need.

The most cognitively natural path is not always the most effective for information seeking.
Participants frequently encountered dead-ends when habitual strategies overshadowed their actual information needs (\textbf{C3 — Overreliance on Cognitive Shortcuts}).
For instance, P1 would always \qt{first ask in Chinese [their native language] and see what kind of answers [they] get}, since it was easiest for her to read and write.
This strategy, however, often failed to satisfy her information needs, requiring additional effort to switch to another language such as English.
\fixme{The tacit \pqt{conversion process}{P10} of mapping concepts across languages can be mentally burdensome, hindering idea connection and discouraging switching to the better language when users’ initial opportunistic language choice is insufficient.}

\textbf{\Practical{} Factors.}
\fixme{Our study also identified aspects of language scent related to the costs and effort required for cross-language interaction within the system or infrastructure.}
Currently, participants are often deterred from searching in multiple languages because \pqt{information from different languages are captured in different silos}{P9}, making transitions between them expensive (\textbf{C4 - Inadequate Cross-Language Integration}).
Part of the problem is that there is currently no \pqt{unified entry point to search for information}{P10} across languages. Users therefore need to \pqt{manually add a middle step in between}{P1} where they \pqt{translate and redo [their search] to build up for seeking information in the other silos}{P9}, which can be tedious and time-consuming.
This finding echoes prior work on query reformulation ~\cite{fuQueryReformulationPatterns2017, fu2018MixedLanguageQueries, albarillo2018InformationCodeSwitchingStudy}.

Even after successfully obtaining information in their desired language(s), participants found \pqt{processing them [the information] tiring}{P6} because \pqt{there's too much information [...] and a lot of it is redundant and useless}{P2} (\textbf{C5 – Information Overload}). 
The challenge of \pqt{consolidating the information}{P5} went beyond simple summarization and extended to the process of triangulation.
Participants particularly desired \pqt{seeing the similarities and differences between sources in different languages}{P7} in order to \pqt{make it clear the stance and positionality of each}{P1}.
This requires extensive back-and-forth conversion and cross-referencing across languages, adding significant effort.

\subsection{Design Guidelines}
\label{sec:design}

\rev{Overall, the previous section shows that the language scent of a language subspace depends both on the \textbf{information it contains} (epistemic factors) and on \textbf{users' ability to extract that information} (interpretative and practical factors).
Thus, a system for language-scent-guided cross-language information navigation should provide cues that support users in assessing both aspects.
}
We derive four \textbf{d}esign \textbf{g}uidelines (\textbf{DG\#}) to address the identified challenges.
\begin{enumerate}[label=\textbf{[DG\arabic*]}]
    \item Provide \rev{proximal cues} that support evidence-based \rev{preview} of \rev{the} information \rev{present} across different languages. (\textbf{C1})
    \item Expose users to multiple languages to reduce overreliance on cognitive shortcuts and activate cross-linguistic {thinking}. (\textbf{C3}) 
    \item \rev{Increase awareness of one's own cross-language search patterns to help users adapt and improve their strategies. (\textbf{C2})}
    \item Fulfill these guidelines while minimizing unnecessary effort and operational overhead for users. (\textbf{C4, C5})
\end{enumerate}

\section{\sys{}: \rev{Language-Scent-Guided Navigation}}

\begin{figure}[t]
    \centering
    \includegraphics[width = 1\linewidth]{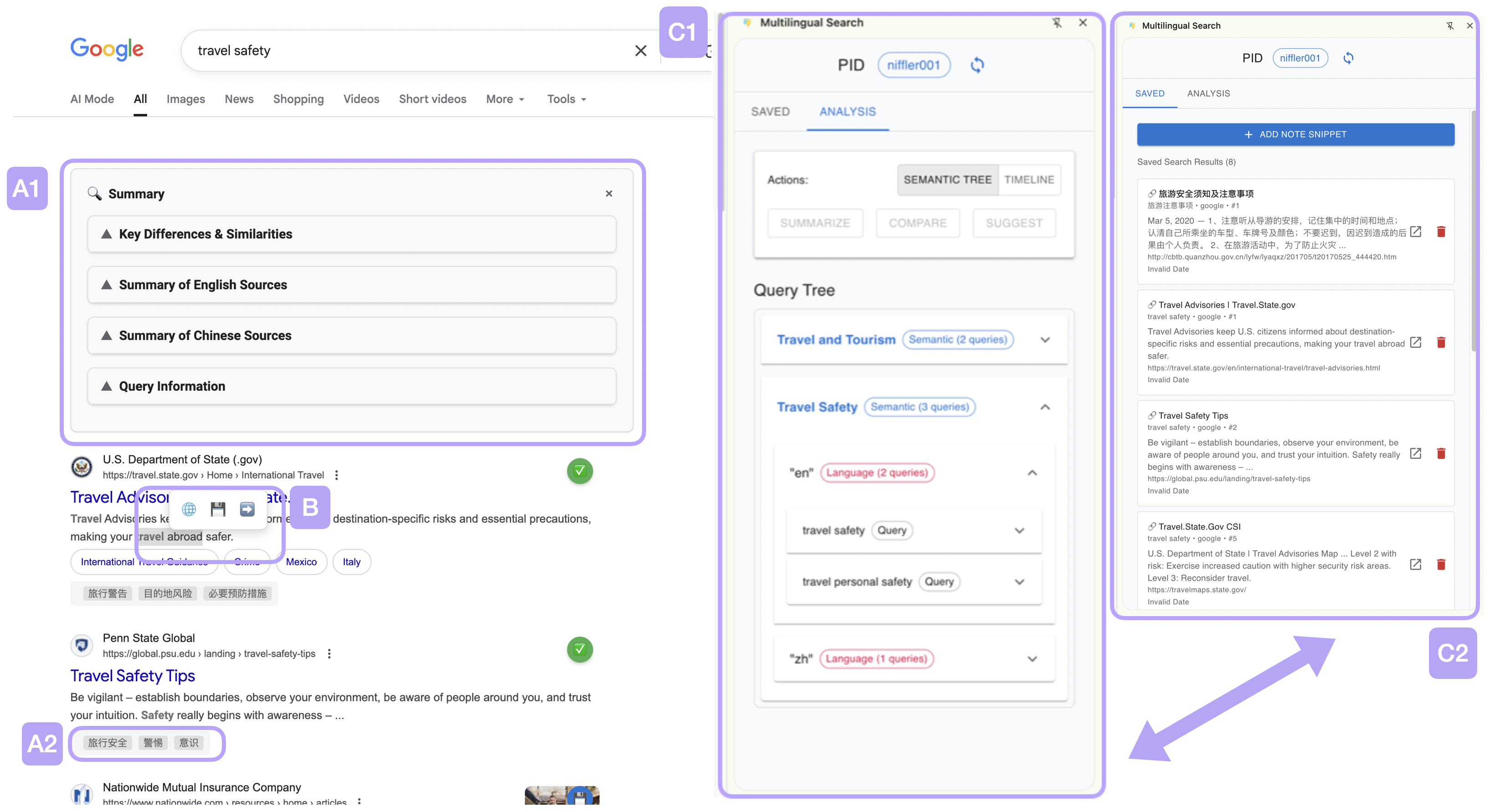}
    \caption{System Overview. \sys{} consists of [A] a search results page enhancement of [A1] Comparative Summary and Query Information, and [A2] Cross-Lingual Keywords, [B] in-situ tools, and [C] a side panel with tabs [C1] analysis and [C2] saved content.}
    \label{fig:sys}
\end{figure}

\rev{Based on our design guidelines, we developed \sys{}, a multilingual information seeking system enabling language-scent-guided navigation.
\sys{} helps users better gauge the value of each language subspace through proximal cues about:
\begin{itemize}
    \item the information available in each language (\textbf{DG1}; Section \ref{sec:DG1})
    \item how their existing knowledge and language proficiency align with the information being sought (\textbf{DG2}; Section \ref{sec:DG2})
    \item whether their language-related search patterns are yielding proportionate value (\textbf{DG3}; Section \ref{sec:DG3})
\end{itemize}
To prevent the practical cost of switching languages from diminishing other factors of language scent (\textbf{DG4}), \sys{} consolidates cross-language search into a single entry point, removing the need to manually repeat queries in each language.
\sys{} is further designed to minimize the cognitive costs (\textbf{DG4}) of seeing information from multiple languages by following an overview-first, detail-on-demand approach where users see high-level summaries and cues by default and can drill down as needed.
As \textbf{DG4} informs the design of all features, it is integrated into other sections and not presented separately.
}

\begin{figure*}[t]
    \centering
    \includegraphics[width = 0.7\linewidth]{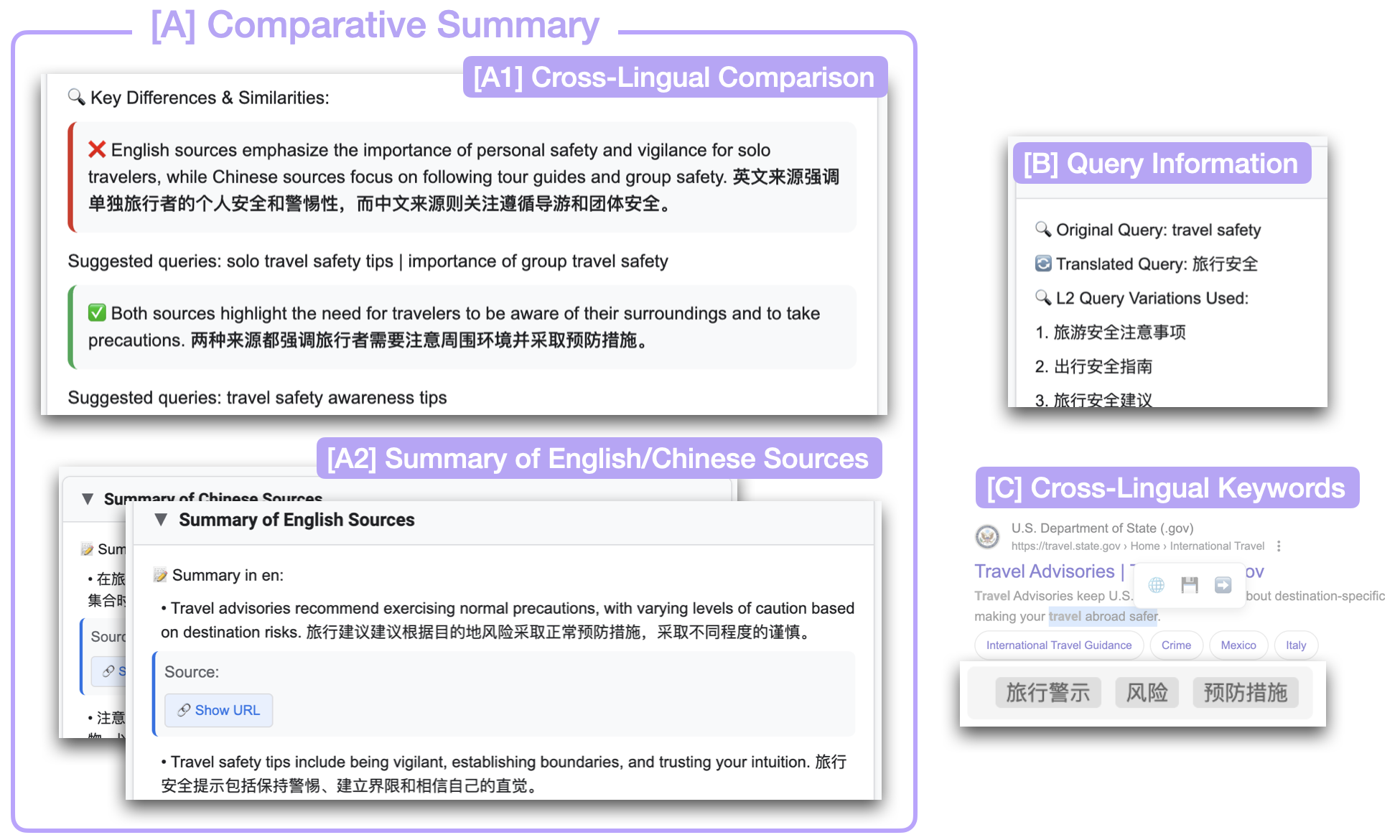}
    \caption{On the search results page, the [A] Comparative Summary consists of [A1] Cross-Lingual Comparison summarizing key similarities and differences between the two languages, and [A2] Summaries of English sources and Chinese sources, respectively. [B] Query Information is provided as background information to complement it. Each search result is decorated with [C] Cross-Lingual keywords.}
    \label{fig:sysSummary}
\end{figure*}

\begin{figure}[t]
    \centering
    \includegraphics[width = 0.9\linewidth]{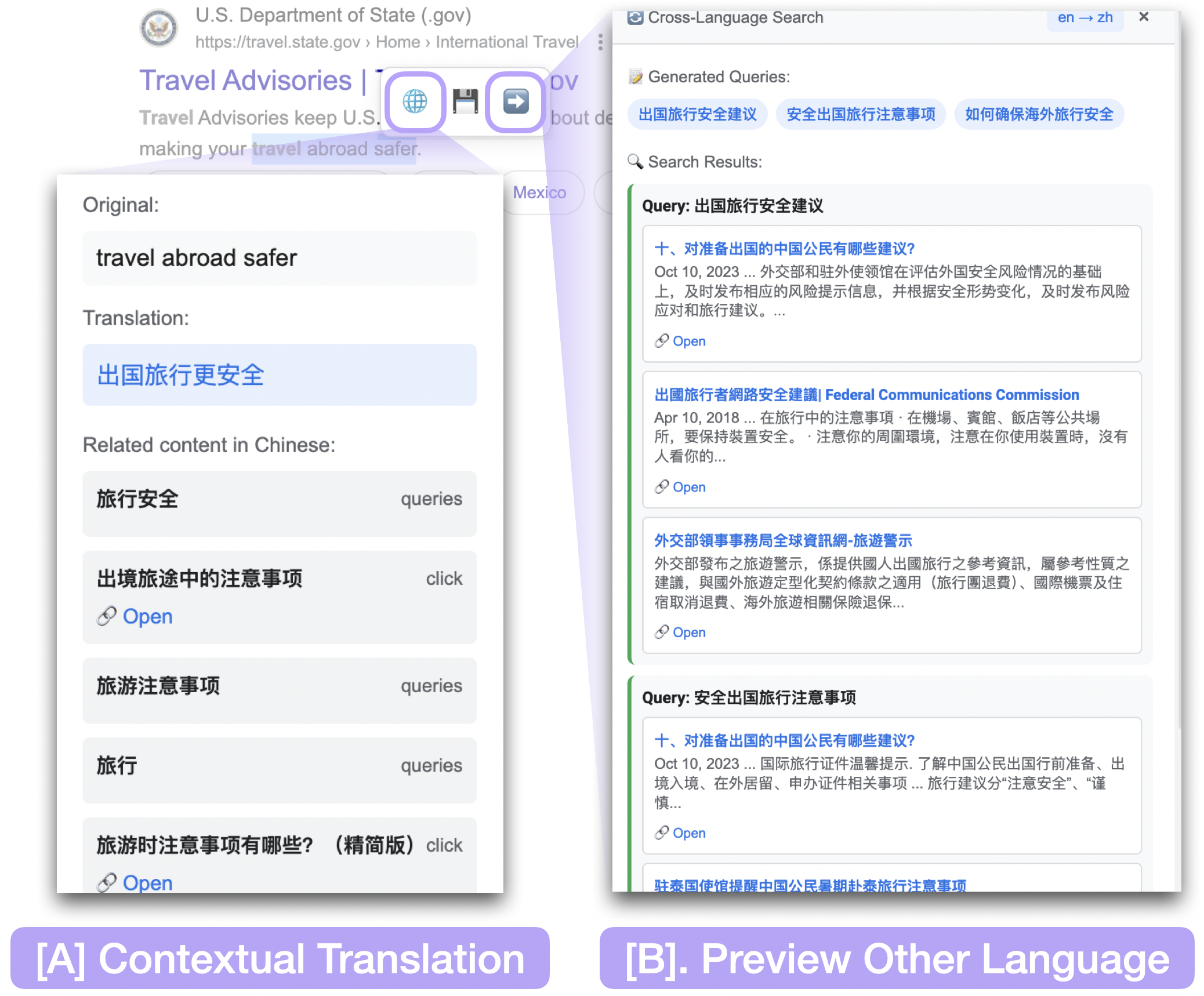}
    \caption{The tooltip opens when text is selected. Users can choose to see [A] Contextual Translation of the text, connecting to relevant user search history, or [B] Preview of Other Language, with suggested queries and sources in the other languages.}
    \label{fig:sysTooltip}
\end{figure}

\subsection{System Overview}
\sys{} comprises three main components (Fig.~\ref{fig:sys}): (1) a search results page augmentation for forming search strategies (Fig.~\ref{fig:sys}[A1, A2]), (2) tooltips for connecting across languages (Fig.~\ref{fig:sys}B), and (3) an analysis side panel for reflecting on search strategies (Fig.~\ref{fig:sys}[C1]) and note-taking (Fig.~\ref{fig:sys}[C2]).
We use the user's queries, clicks, saved content, and notes as a proxy of their search activity in the backend~\cite{joachims2017AccuratelyInterpretingClickthrough}. 
In particular, we treat queries as the smallest natural unit of search activity and organize clicks, saved content, and notes around them.
Users can save queries and their corresponding search results or webpage snippets when using \sys{}, or create custom \textbf{notes} in the side panel.
\rev{\sys{} is not personalized to avoid narrowing results and undermining the system's exploratory goal.}

We implemented \sys{} as a Google Chrome extension using the Chrome Extension API.
The front-end was developed in TypeScript with React and Material UI, while the back-end was built in Python with FastAPI for handling API calls.
Data were stored in Firebase and indexed and searched with TypeSense.
We used the Google Search API to retrieve relevant sources and OpenAI's GPT-4o API for summarization and query generation.

\subsection{\rev{Epistemic Cues} (DG1)}
\label{sec:DG1}
\sys{} helps users gauge the informational value of languages, thereby strengthening language scent.
\subsubsection{Always-On Overview}
For every search, users can view a \textbf{Comparative Summary} (Figure \ref{fig:sysSummary}A) of English and Chinese sources, helping to reduce information overload (\textbf{DG4}).
It includes a {Cross-Lingual Comparison}, which highlights similarities and differences between the two languages and provides suggested queries to further explore the points of comparison (Figure \ref{fig:sysSummary}[A1]).
{Summaries of Sources} in each language are also provided, summarizing the key points with linked sources (Figure \ref{fig:sysSummary}[A2]).
The \textbf{Query Information} section shows the queries used to retrieve the sources underlying the summaries (Figure \ref{fig:sysSummary}B).

To obtain information from {both languages}, we followed a pipeline of translating and rewriting the query in the other language \cite{hull1996QueryingLanguagesDictionarybaseda, sheridan1996ExperimentsMultilingualInformation}, retrieving relevant sources from search engine result pages \cite{palani2021CoNotateSuggestingQueries, palani2022InterWeavePresentingSearcha}, clustering \cite{wang2011IntegratingDocumentClustering}, and summarizing \cite{christensen2014HierarchicalSummarizationScaling}, following common information retrieval approaches \cite{zhu2024LargeLanguageModels}.
\rev{The summaries and keywords are generated using OpenAI's GPT-4o and ranked by relevance to the query.
A technical evaluation (Appendix \ref{app:techEval}) using queries and results from the lab study (Section \ref{sec:lab}) suggests the system's outputs are generally accurate and relevant.
}

\subsubsection{In-Situ Tool}
Information seeking is a serendipitous process.
To support this, the \textbf{Preview Other Language} function (Figure \ref{fig:sysTooltip}B) lets users view content from another language without leaving the current context \fixme{by selecting text and clicking the tooltip button}.
It provides suggested queries and relevant sources in the other language, allowing users to assess whether switching languages is worthwhile while cross-referencing with information in the current language, without needing to context-switch and do a full new search (\textbf{DG4}).

\subsection{\rev{Interpretative Cues} (DG2)}
\label{sec:DG2}

\rev{\sys{} helps users gauge how effectively they can extract information in a language for the current topic, which depends on factors like contextualized language proficiency, by showing cues in the other language to prime users' latent knowledge.}

\subsubsection{Always-On Cues}
The comparative summary and search activity analysis functions are displayed in both languages.
For search results, translating each of them entirely could create overload and cognitive stress (\textbf{DG4}), and is not space efficient.
Instead, \textbf{Cross-lingual Keywords} (Figure \ref{fig:sysSummary}C) summarize the content of individual sources in the other language.

\subsubsection{In-Situ Tool}
To address ad-hoc needs to connect languages, the \textbf{Contextual Translation} function (Figure \ref{fig:sysTooltip}A) translates selected text into the other language and shows relevant items from the user's search activity in the other language, i.e. queries, clicks, saved content, notes.
The relevant items are retrieved using both query translation \cite{hull1996QueryingLanguagesDictionarybaseda} and embedding-based \cite{vulic2015MonolingualCrossLingualInformation} approaches.
This helps users connect knowledge and intermediate search results across their two languages without having to manually sift through past records (\textbf{DG4}).

\subsection{\rev{Reflective Cues} (DG3)}
\label{sec:DG3}
\begin{figure*}[t]
    \centering
    \includegraphics[width = 0.7\linewidth]{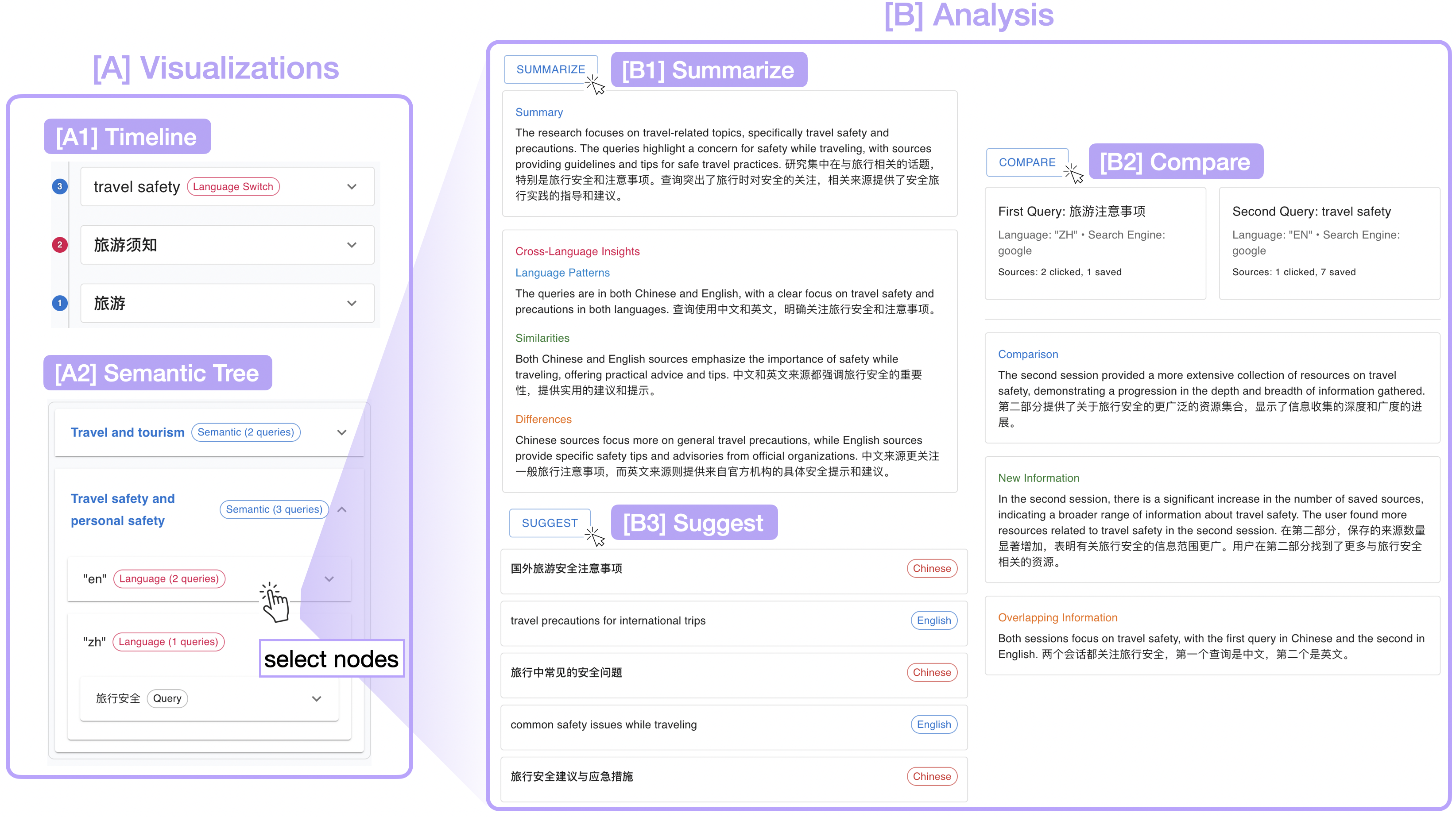}
    \caption{The side panel analysis tab provides [A] language-centred visualizations of user search activity and [B] analysis functions to help them consolidate information and evaluate past strategies.
    There are two visualizations, the [A1] Timeline highlighting language switching patterns and the [A2] Semantic Tree focusing on high-level concepts. Users can select nodes on the visualization for analysis, being able to [B1] Summarize, [B2] Compare, or [B3] Suggest queries.}
    \label{fig:sysVis}
\end{figure*}

\sys{} helps users \rev{examine and refine their past language subspace navigation decisions}, by allowing them to view and analyze their search activity (Figure \ref{fig:sysVis}).
\rev{This gives users a bird's-eye view of their decisions and outcomes across a session, helping them gauge whether their language choices were paying off.}

\subsubsection{Visualizations}
We introduced two language-centred visualizations of users’ search activity (Figure \ref{fig:sysVis}A).
The \textbf{Semantic Tree} (Figure \ref{fig:sysVis}[A2]) organizes past searches first by subject matter and then by language, allowing users to see their language use for different topics.
The \textbf{Timeline} (Figure \ref{fig:sysVis}[A1]) presents searches chronologically, highlighting points of language switching.
In both representations, users can click on query nodes to expand them and view the sources and notes related to them.

\subsubsection{Analysis Functions}
We also provide three complementary {analysis} functions (Figure \ref{fig:sysVis}B) to help users interpret their search history and the information gathered with less overhead (\textbf{DG4}).
The \textbf{Summarize} function synthesizes the content of selected nodes, offering an overview and cross-language comparison of sources (Figure \ref{fig:sysVis}[B1]).
The \textbf{Compare} function dives one step deeper by showing the marginal benefit of a later-selected query relative to an earlier one by identifying new versus overlapping information (Figure \ref{fig:sysVis}[B2]), for example allowing users to better evaluate the value of switching languages or remaining in a language.
The \textbf{Suggest} function facilitates further exploration and expansion by recommending additional queries to extend the selected nodes in both languages (Figure \ref{fig:sysVis}[B3]).

\section{Lab Study}
\label{sec:lab}
We conducted a lab study of \sys{} to understand how its features augmenting language scent influence users' multilingual information seeking.

\subsection{Method}

We followed a within-subjects design with two conditions (\sys{} and \base{}) and two tasks.
The conditions and tasks were fully Latin-square balanced to mitigate potential order effects.
%Below, we explain our rationale for baseline and task selection.  

\begin{figure}[t]
    \centering
    \includegraphics[width = 1\linewidth]{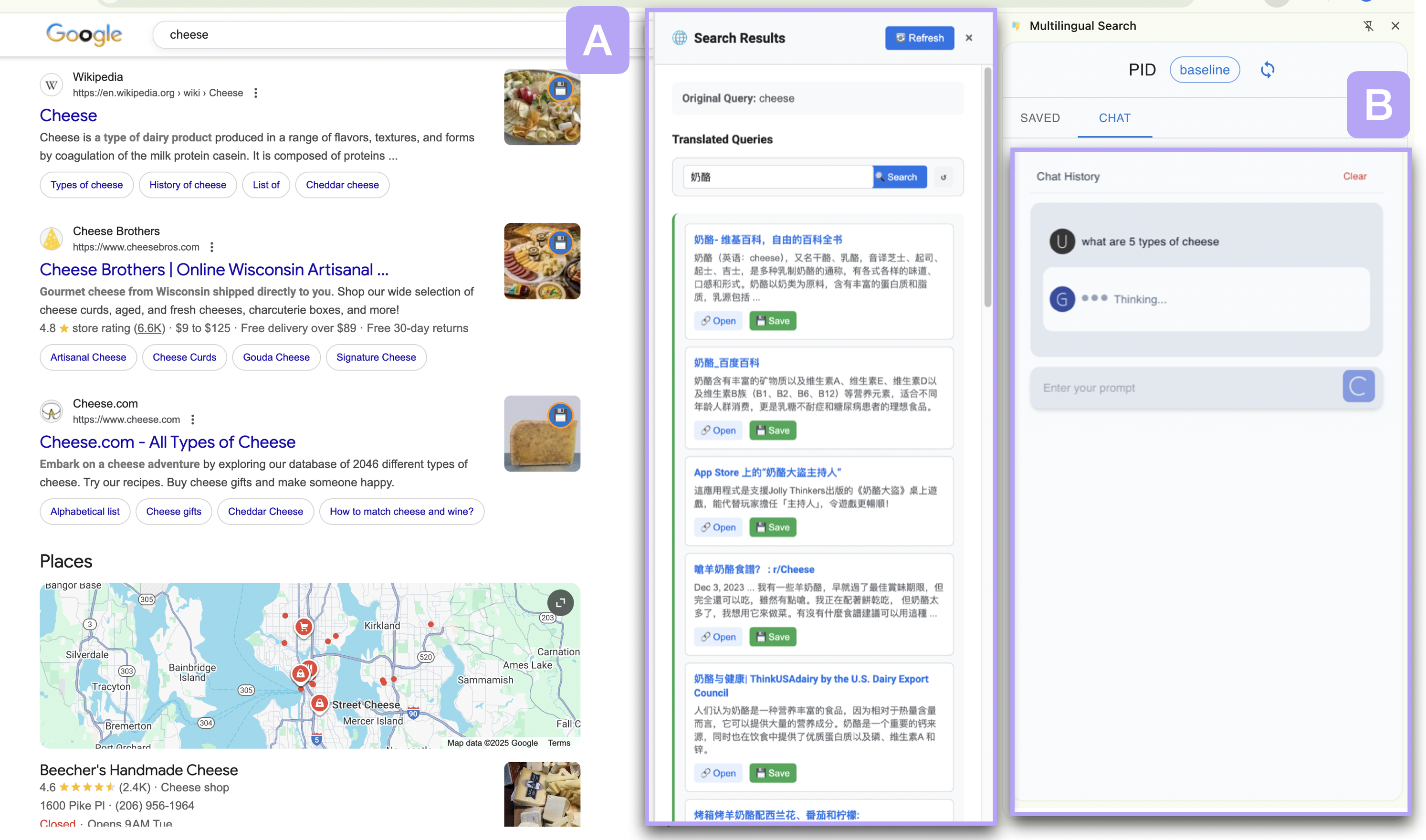}
    \caption{The \base{} consists of [A] Parallel Search Panel and [B] AI Chat Panel.}
    \label{fig:baseline}
\end{figure}

\subsubsection{Conditions}
The \base{} condition (Figure \ref{fig:baseline}) consisted of a parallel search interface and an AI chat panel, representing an enhanced version of status quo tools \cite{mayerhofer2025BlendingQueriesConversations}.
We did not directly compare against existing tools, as they operate in only one language, which would make the baseline inherently less information-dense than \sys{}.

While no commercially available tool currently supports multilingual search, prior work has explored interfaces that display results in multiple languages \cite{steichenSupportingModernPolyglot2015, chuTranSearchMultilingualSearch2017}.
From this work, we adopted the panel design for our \base{} condition that was shown to be most preferred in \citet{chuTranSearchMultilingualSearch2017}.
The AI chat panel was implemented using OpenAI’s API to ensure consistency and avoid variability from participants using different models.
To enable a fair comparison, \base{} also includes the same saving and notes functionalities as \sys{}.

\subsubsection{Task}
We created two exploratory information seeking tasks (informed by~\cite{carevic2018InvestigatingExploratorySearch}) that were open-ended, likely to occur in real-world settings, broadly applicable, and designed to minimize bias toward English or Chinese contexts. Participants were asked to collect diverse information on two topics (Appendix \ref{appendix:task}): (1) Career Advice and (2) Food and Restaurant Recommendation in Switzerland (chosen because its official languages do not include English or Chinese \cite{languageSwitzerland}).
Participants were given 20 min for each task, based on prior research on the average duration for conducting exploratory online information seeking \cite{carevic2018InvestigatingExploratorySearch, athukorala2016ExploratorySearchDifferent}. They were also encouraged to take notes on the fly to stay engaged with the task.

\subsubsection{Data Collection and Analysis}
We collected Likert-scale ratings on ten questions (Appendix \ref{app:likert}), \rev{designed around \sys{}'s design guidelines and features to capture participants' perceptions of its ease of use, as well as their experiences forming and reflecting on language scent, following standard practice in HCI system evaluation \cite{liu2024SeleniteScaffoldingOnline, suh2023SensecapeEnablingMultilevelb}.}
We also collected system logs and screen recordings from the sessions.
For participants’ navigational behaviour during information seeking, logs of user queries were used as a proxy \cite{gwizdka2006WhatCanSearching, mathassan2005AssociatingSearchNavigation}.
The measures we examined are: \textit{number of queries}, number of \textit{language switches}, number of consecutive queries with each language (\textit{language span}), and distribution of queries across languages (\textit{language balance}) through Shannon's entropy \cite{shannon1948MathematicalTheoryCommunication, lin1991DivergenceMeasuresBased} (details in Appendix \ref{appendix:logMeasure}).
For a proxy of the relevant information participants gathered, we used logs of the sources participants clicked \cite{joachims2017AccuratelyInterpretingClickthrough}, or saved or took notes on \cite{zhengDiscipLinkUnfoldingInterdisciplinary2024a}.
Two coders independently coded the topic coverage of these sources for each participant–task, blind to condition, following the iterative procedure in \cite{richardsPracticalGuideCollaborative2018} (details in Appendix \ref{appendix:logCoding}).
Finally, for qualitative feedback, we open-coded the interview transcripts to gain a systematic and structured understanding \cite{braunUsingThematicAnalysis2006a} of how users perceived and interacted with \sys{}.
\rev{Since our goal was to characterize the breadth of participant responses rather than build a formal model, we focused on open coding to capture distinct observations without imposing higher-level code structure through axial coding.
Screen and audio recording were used when context was needed or the transcripts were ambiguous.
}

\subsubsection{Participants}
We recruited 16 participants (14 women, 2 men; mean age $= 25 \pm 2.18$) via social media and snowball sampling.
The sample size was determined based on an a-priori power analysis ($\alpha = 0.05$) for detecting a medium (Cohen's $d = 0.75$) effect size \cite{cohen1992StatisticalPowerAnalysis, ortloff2025SmallMediumLarge}.
All participants were native speakers of Chinese and at least independent users of English according to the Common European Framework of Reference for Languages (CEFR) \cite{CEFRLevelsCommona}.
\rev{All participants lived in Mainland China until at least age 14 ($M = 18.7, SD = 3.4$), and at the time of the study, resided in an English-speaking country (10 in Canada, 5 in the United States, 1 in the United Kingdom).}
All participants regularly engaged in online information seeking (mean $= 2.76 \pm 0.66$ sessions a day).
\rev{Additionally, on average participants rated their frequency of information seeking in Chinese as $6.5$ and in English as $6.25$ on a 7-point Likert scale ($1 = \text{Very Rarely}$, $7 = \text{Very Frequently}$).}

\subsubsection{Procedure}
We conducted the study remotely via Zoom, starting by obtaining participant consent.
Before proceeding to the tasks, participants were asked to install a Chrome extension containing both \sys{} and \base{}.
Before each task, we explained the system to participants and provided time for them to familiarize themselves with it as needed.
Each task session lasted 20 minutes, and participants were encouraged to think aloud \cite{ThinkingAloud1, charters2003UseThinkaloudMethods}.
After each task, participants completed a questionnaire with Likert-scale items assessing their experience.
The study lasted approximately 90 minutes, and participants were compensated US \$25.
Task sessions and interviews were audio- and screen-recorded, then transcribed verbatim.
Their interactions with the system were logged.
The study protocol was approved by our institution's ethics review board.

\subsection{Lab Study Findings}
\label{sec:findings}

\begin{figure*}[t]
    \centering
    \includegraphics[width = 0.9\linewidth]{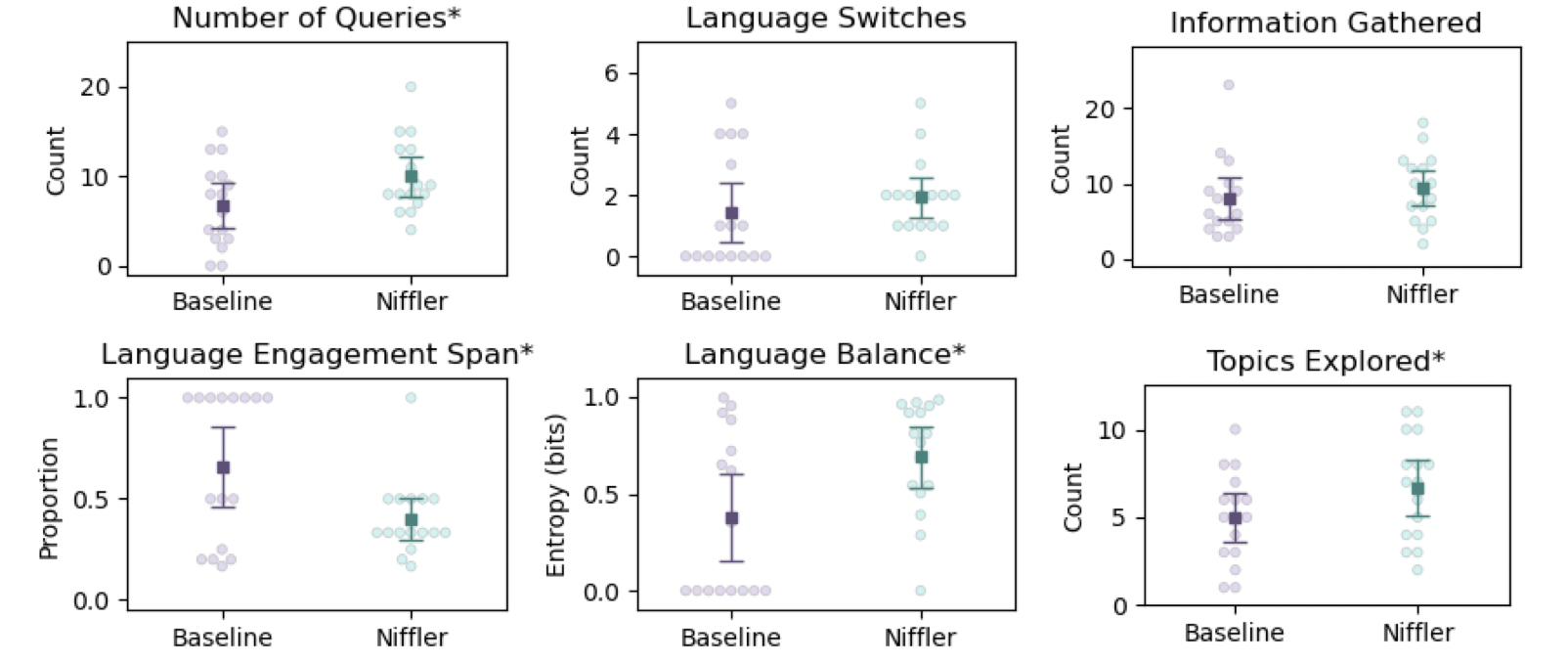}
    \caption{Swarm Plots for Analysis on Information Seeking Process and Outcomes. Square mark indicates mean and error bars are 95\% confidence intervals. Asterisk (*) indicates statistical significance. \textbf{Process}: In \sys{}, participants searched significantly more, stayed in the same language for significantly shorter duration, and the language distribution of their queries was significantly more balanced, compared to \base{}. There was no significant difference in the number of times they switched the query language. \textbf{Outcome}: There was no significant difference in the amount of information gathered, but participants explored significantly more topics with \sys{} compared to \base{}.}
    \label{fig:logStat}
\end{figure*}

We report our qualitative (from participant interviews) and quantitative (from the questionnaire and system log data) findings below.
For quantitative data, we applied the Wilcoxon signed-rank test to calculate statistical significance, as it is a non-parametric method that makes no assumptions about the underlying distribution \cite{bridge1999IncreasingPhysiciansAwareness}.
We applied the Benjamini-Hochberg correction to the Likert-scale questionnaire to control for family-wise false discovery rate.
Full statistics can be found in Table \ref{tab:likert} for Likert-scale items and Table \ref{tab:log} for system logs (Appendix \ref{app:stats}).
Here, we summarize the key results, reporting the mean ($M$), $p$-values, and effect size ($r$).

\subsubsection{\rev{{Proximal} cues augment language scent}}
From the Likert-scale items, it was significantly easier to identify similarities and differences across languages in \sys{} compared to in \base{} ($M_{\text{\sys{}}} = 1.88 < 3.56 = M_{\text{\base{}}}; p = 0.044, r = 0.829$; lower is better).
No significant difference was found in the other two Likert items on forming language scent (Appendix \ref{app:likert}.3).

Participants had amplified awareness of cross-language differences in \sys{}:
\begin{quote}
    \pqt{I feel \sys{} is a bit beyond my expectations --- it really exceeded what I imagined. I didn’t expect that there would be such a big difference between the Chinese and English web.}{P5}
\end{quote}
The surprising insights tend to \pqt{provide inspiration}{P2} and \pqt{trigger [participant's] interest in checking more}{P1}, even if they \pqt{didn't think to search in this [a particular] language at first}{P14}.
This motivated them \pqt{to think about a problem from more diverse perspectives and to gather information from more diverse angles}{P3}.
While \base{} had some nudging effect as well, participants \pqt{didn't pay attention or notice as much}{P2} because \pqt{it doesn't clearly distinguish between results and [...] doesn't help understand general trends}{P9} across languages.

On the \textbf{epistemic} side, participants drew on the similarities and differences across languages in the \textit{Comparative Summary} to guide their search strategies, and to extract useful information for getting started.
Similarities were perceived as \pqt{more common and widely recognized, rather than culturally specific}{P13}, which participants either trusted more readily or treated as indicators of importance to investigate further, depending on their goals.
Differences, on the other hand, were used to \pqt{identify variations in cultural emphases}{P12} and broaden perspectives.
Participants scrutinized differences more carefully than similarities to assess their underlying causes and relevance before deciding whether to accept the information, particularly when the differences were \pqt{unexpected or didn't align with [their] intuition}{P4}.

\sys{}’s design of showing summaries and content in both languages also helped {amplify} the interpretative factor of language scent by \pqt{allowing [participants] to better understand the content}{P8} through cross-referencing.
In particular, the \textit{cross-lingual keywords} were valuable since \pqt{having tags in a different language felt like having two distinct pieces of information [...] if the tags were in the same language, it probably won't help as much}{P2}.
\textit{Cross-lingual keywords} also had a subtle nudging effect \pqt{connecting to [participants'] knowledge and helps [them] gauge if checking what the other language says is better}{P3}.
{These features help participants recognize their preferences for working with specific topics or information through contextual cues.}

\subsubsection{{Augmented language scent facilitates formation of search strategies}}
The system log of user queries (Figure \ref{fig:logStat}) showed that users formulated significantly more queries ($M_{\text{\sys{}}} = 10.0 > 6.75 = M_{\text{\base{}}}; p = 0.041, r = 0.549$) when using \sys{}.
\rev{\sys{}'s workflow also helps users adjust their search strategies more flexibly and promptly in response to evolving information needs than \base{} does.}
{There was no significant difference in the average number of times} participants switched the query language {across the two conditions}.
However, after switching, they stayed within the same language for significantly fewer consecutive queries than \base{} (normalized; $M_{\text{\sys{}}} = 0.403 < 0.657 = M_{\text{\base{}}}; \allowbreak p = 0.017, r = 0.642$), i.e. switched more quickly.
These align with participant observations that \sys{} enabled them to form \pqt{more fine-grained and effective strategies}{P6}, whereas in \base{} they tended to \pqt{stick to one or completely switch languages mid-way}{P6}, in a more opportunistic manner.
Perhaps a consequence of this, participants issued queries across the two languages significantly more evenly in \sys{} than \base{} (Shannon entropy \cite{shannon1948MathematicalTheoryCommunication, lin1991DivergenceMeasuresBased}; $M_{\text{\sys{}}} = 0.690 > 0.381 = M_{\text{\base{}}}; p = 0.020, r = 0.601$).

\rev{Participants thought that \sys{} led to a more exploratory and flexible workflow compared to \base{}, where they \pqt{branched out}{P11} more.
Participants also \pqt{digged deeper into details}{P16} when using \sys{}.
}
Participants found that whereas they had to primarily \pqt{rely on [their] own intuition}{P6} in \base{}, \sys{} allowed them to be more \pqt{targeted}{P6} and \pqt{intentional}{P16} when switching languages, providing \pqt{a clear idea and some inspiration for next steps}{P4}.
For example, P9 described how \sys{} \rev{helped} them efficiently form a search strategy for career advice:
\begin{quote}
    \pqt{When searching for employment goals, the system [\sys{}] helped me realize the results in Chinese and English were different. The Chinese results were mostly related to promotions or government statements, encouraging people to apply for specific jobs. For me, that kind of information wasn’t very useful, so I didn't want to spend time searching in Chinese. Instead, I mainly looked at the English results, which helped me avoid spending extra time on repetitive searches.}{P9}
\end{quote}
\sys{} also supported participants in flexibly adapting and refining their strategies.
Continuing the previous example, \sys{} later helped P9 realize that unlike for employment goals, for information on \pqt{time management, [they] actually prefer Chinese because for English it's more websites from specific universities, with timelines targeted towards their own students, whereas in Chinese it's more general}{P9}.

\subsubsection{{Augmented language scent facilitates reflection on search strategies}}
While we did not find significance for the two Likert-scale items on facilitating reflection on language scent (Appendix \ref{app:likert}.3), this may be due to the nature and duration of the task.
Participants reported that while \pqt{the system [\sys{}] supports reflection … and [they] have some thoughts during the process}{P7}, they \pqt{feel like [they] haven’t reached the stage of reflecting yet}{P7}, and therefore many did not engage substantially with the reflection features.
Participants did note that \sys{} encouraged meta-level reflection on their language scent beyond the current session.
In addition to evaluating their language scent through the search outcomes, participants were most interested in whether their engagement with different topics was skewed toward a single language.
Participants mentioned that \sys{} \pqt{helped [them] realize [them] always subconsciously choosing a specific language, something that [they] might not otherwise have noticed}{P6}, for example using the conceptual groupings in the \textit{Semantic Tree} to understand language distribution across topics.
Participants did not regard uneven use of the two languages as intrinsically positive or negative.
{Instead, it prompted them to reflect on cross-language differences in context of the different topics, and, when deemed substantial, to \pqt{adapt [their] strategies not just now but also in the future}{P6}.

\subsubsection{Augmented language scent leads to more diverse information gathered.}
\rev{Because our study tasks elicited exploratory behaviour by asking participants to gather diverse information, we use topic coverage (Figure~\ref{fig:logStat}) as a measure of task success.}
There was no significant difference between the number of relevant sources participants found across \base{} ($M_{\text{\base{}}} = 8.13$) and \sys{} ($M_{\text{\sys{}}} = 9.44$).
However, with \sys{}, participants were able to explore significantly more topics ($M_{\text{\sys{}}} = 6.69 > 5.00 = M_{\text{\base{}}}; p = 0.032, r = 0.597$), compared to \base{}.
\rev{To further characterize the topics explored, we analyzed the distribution and nature of topics covered across conditions: participants tended toward broader topics in Baseline (e.g., "career advice"), while in \sys{} they covered a greater number of more fine-grained subtopics (e.g., "graduate student career advice").
}
Relatedly, in their interviews, participants also found that searching in multiple languages \pqt{gives [them] interesting perspectives}{P16} and \pqt{reveals things [they] wouldn't be aware of otherwise}{P3}.
Since participants gained more diverse information while accessing a similar number of sources, \rev{our findings suggest that \sys{} may better support exploration of multilingual information.}

\subsubsection{Usability of \sys{}}
For the five self-reported Likert items on overall impression and ease of use (Appendix \ref{app:likert}.1-2), no significant differences were observed; these smaller-than-expected effects may partly reflect \sys{}’s \pqt{steeper learning curve}{P10}.
Indeed, system logs show that only 8 of 16 participants ever switched languages in \base{}, compared to 15 in \sys{}, potentially suggesting that operational barriers of switching languages were lower in \sys{}, even if participants did not perceive this subjectively.
In interviews, participants noted that once they became familiar with the features, \sys{}’s \pqt{tool assistance encouraged [them] to do multilingual information seeking}{P6}.
Compared to their status quo and the \base{} workflow, participants described a \pqt{trade-off between time and effort versus the result}{P1}, suggesting that \sys{} lowered the effort required to explore multiple languages.
Participants liked that \pqt{it [\sys{}] offers high-level insights}{P1} such as \pqt{the key differences between Chinese and English sources}{P8}, and does not \pqt{require a lot of effort to compare or process the information by [oneself]}{P1}.

\subsubsection{Envisioned Real-Life Use}
Our findings indicate that \sys{} may be the most helpful for exploratory tasks, where the goal is to \pqt{hear as many voices and opinions as possible}{P5}, and less useful for transactional tasks, where a single clear answer can typically be obtained through one search.
Exploration may be desirable \pqt{when [they] care a lot about the truthfulness of the information}{P4} (high-stakes) or when \pqt{they are learning about a completely new topic}{P3} (limited prior knowledge).
There was no clear pattern in the topics where \sys{} was expected to be helpful.
Scenarios mentioned range from everyday tasks like \pqt{buying cars}{P5} and \pqt{trip planning}{P1}, to more serious ones like \pqt{medical advice}{P10} and broader literature exposure in \pqt{research}{P1}.
Participants envisioned \sys{} as an always-on support as \pqt{it doesn't hurt to have more information, or someone comparing information from different sources for you}{P1}, especially since it's \pqt{hard to predict when there are differences across languages}{P13} and language scent \pqt{evolves over time, adapting to [their] search outcomes}{P6}.
Additionally, participants explained that they could easily disregard irrelevant information when it is not applicable.
% exploratory but not transactional, explore-exploit dilemma in information seeking \cite{athukorala2016RelevanceAdaptingExploration}.

\section{Discussion}

\subsection{{\rev{Views} of Multilingual Information}}

Language scent is shaped both by users’ prior \rev{expectation} of multilingual information and by the environmental cues they encounter when interacting with the system.
\sys{} surfaces previously obscured language scent through explicit system cues, facilitating the refinement of users’ priors and enabling us to study their views of multilingual information as a whole.
\rev{By analyzing the lab study task and interview recordings}, we found that users displayed three main \rev{views} regarding the role of information from other languages, which we explain below.
Note that the user's view is not static but evolves based on their information seeking experience.

\paragraph{\textbf{Multilingual Information as Fallback}}
In this view, the prior expectation of multilingual information is low and users treat information in another language as a fallback, switching only when they are unable to find satisfactory results in their primary language.
For example, P2 represents an extreme case, as they did not switch languages at all during the task, explaining that \pqt{if Chinese [the other language] returns significantly better results than English [the primary language] [...] but [they] don't expect it to be the case}{P2}.
Their low expectation led them to largely ignore the proximal cues provided by \sys{}, leaving their information seeking driven almost entirely by information scent.
Only 3 out of 16 participants had this \rev{view} (P2, P3, P11) at some point, with P3 shifting towards the ''safeguard'' view after using \sys{}.

\paragraph{\textbf{Multilingual Information as Safeguard}}
In this view, the prior expectation of multilingual information is moderate and users treat information in another language as a way to cross-reference and verify information.
Although users also adopt a primary search language in this case, they are more willing to switch languages {than in the ''fallback'' case}.
When the accuracy or impartiality of information is especially important, participants {with} this view would \pqt{check if the languages agree and if [they] missed any perspectives}{P16}.
Otherwise, for easier tasks (e.g., transactional queries) or when they care less, they tend to remain in their primary language, mainly guided by their information scent.
10 out of 16 participants displayed this \rev{view} (P3, P4, P5, P6, P7, P10, P12, P13, P14, P15), with three of them (P5, P6, P12) shifting towards the ``complementary resource'' model after interacting with \sys{}.

\paragraph{\textbf{Multilingual Information as {Complementary Resource}}}
In this view, the prior expectation of multilingual information is high and users treat information in different languages as complementary resources that jointly form a complete understanding.
Participants with this view switched languages freely and frequently, treating both languages as similarly valuable resources.
7 out of 16 participants displayed this \rev{view} (P1, P5, P6, P8, P9, P12, P16), with four \rev{of} them adopting this view after using \sys{}.

\subsection{{Language as a Search Heuristic}}
Our lab study showed that the language scent support in \sys{} encouraged multilingual participants to explore more broadly and consider multiple perspectives by nudging them toward varied sources in different languages.
This suggests that language itself can serve as a heuristic for accessing diverse information, which is beneficial in general, helping with forming informed opinions \cite{bhuiyanNewsCompFacilitatingDiverse2023} and overcoming single-language filter bubbles \cite{quelleLostTranslationMultilingual2023}.
As a heuristic, language has the advantage of being an inherent property of any piece of information, making it scalable and requiring less contextual learning than current heuristics based on platforms or media outlets \cite{wangMediaBiasDetector2025}. 
While our work focuses on cross-lingual information navigation through the lens of language scent, future work could more directly investigate how language can be leveraged as a heuristic for promoting exposure to diverse information.

\subsection{Studying Multilingual Users in HCI}
Multilingual users have long been studied in HCI, with most prior work focusing on limited language competence and technological asymmetry across languages \cite{kimItsMyLanguage2024, choiMultilingualConversationalAgent2023}.
In contrast, our work highlights a complementary and still underexplored approach: focusing on users’ multi-competence.
We show through language scent that \rev{multilingual information seeking is topologically different from monolingual information seeking}, giving rise to new patterns of interaction and distinct cognitive processes.
In this sense, multilingual users are not simply monolingual users replicated across multiple languages, but instead exhibit qualitatively different ways of engaging with information, which is a promising future direction.

\subsection{Limitations and Future Work}
Our work has several limitations and opportunities for future work. 

\textbf{\rev{Applicability across Languages}.}
We recruited people who are fluent in English and Chinese \rev{as a case study of} the broader multilingual population.
Although this is in line with conventional HCI practices \cite{gaoTakingLanguageDetour2022, bawaMultilingualUsersPrefer2020, choiMultilingualConversationalAgent2023} and certain aspects of multilingualism are considered universal regardless of the exact languages spoken \cite{grosjeanBilingualMonolingualLanguage2012, grosjeanBilingualIndividual1997, greenIdentificationCommonalitiesDifferent2023, adjemianNatureInterlanguageSystems1976a}, our empirical findings may not generalize to all configurations of multilinguality, \rev{potentially influenced by the exact languages involved, the linguistic distance \cite{chiswickLinguisticDistanceQuantitative2005} between the languages (e.g., whether they belong to the same language family), settings involving more than two languages, or dialectal variations.}

\rev{While empirical findings may not generalize, language scent is intended as a broadly applicable conceptual lens for explaining multilingual search behaviours. For example, a dialect can serve as a 'distinct search language' when its language scent is high enough. Language scent also extends naturally to three or more languages, involving more choices but the same underlying mechanism. Overall, following the precedent of foundational theories like IFT, we offer a general analytical lens grounded in a case study, leaving other language configurations to future work.}

% study
\textbf{\rev{Study Design and Analysis}.}
We only evaluated \sys{} in a lab environment using two 20-minute tasks \rev{designed to elicit exploratory behaviour}. Although these tasks were designed to be broadly applicable, their limited type and duration inevitably capture only a subset of search behaviours.
\rev{Moreover, their exploratory focus prevents us from assessing downstream outcomes, such as whether exposure to more diverse information or broader exploration leads to greater informedness or better decision-making.}
Future work can address these limitations in two directions: (1) deploying \sys{} over a longer period to understand its utility in more organic settings and investigate long-term behavioural changes, \rev{and (2) designing tasks that more directly evaluate downstream outcomes.}
We also chose a baseline that surfaces information from multiple languages but does not explicitly design for language scent, allowing us to investigate which interface designs best support multilingual information seeking. For future work, monolingual or localized baselines (e.g., Baidu for Chinese) could further isolate the effect of access to multiple languages itself.

\textbf{\rev{Systems for Language-Scent-Guided Navigation}.}
\rev{\sys{} is a first attempt at an information seeking system that enables information navigation guided by language scent.
\sys{} was developed as a generic information seeking tool, as we did not want to bias the design toward a specific application scenario prematurely.
Lab study results later revealed that \sys{} may be better suited to exploratory search, where there is no single answer and encountering different perspectives matters, than to general-purpose search.
Future work could optimize the system for this exploratory task model \cite{ahnPersonalizedWebExploration2008}.
Additionally, \sys{} is designed as an augmentation of traditional web search.
In LLM-mediated search, synthesized responses may suppress source-language cues, and navigation itself may become less central to the search process. This suggests a possible shift in how language scent is operationalized --- from helping users identify where to search, to helping them steer generation to adequately cover relevant languages --- which we leave for future work.
}
Another natural extension is scaling language-scent-guided navigation beyond two languages. This raises both interface-level questions, such as how to design proximal cues that effectively capture information from many languages, and algo\-rithm-level questions, such as how to systematically and scalably incorporate localized engines (e.g., Baidu for Chinese) to optimize search results across languages.

\section{Conclusion}

\rev{In this paper, we introduced language scent --- the perceived value of searching in a particular language --- to explain a decision-layer in multilingual information seeking that IFT does not account for: how users choose which language subspace to search in, prior to navigating between encountered sources.}
Our formative study with 10 English-Chinese bilinguals provides empirical grounding, identifying the factors shaping language scent and the challenges users face.
Based on design guidelines derived from this study, we implemented \sys{}, a \rev{system instantiating language-scent-guided navigation by providing proximal cues.}
A lab study with 16 multilingual participants showed that \sys{} facilitates more exploratory search and more diverse information gathered.
Together, these contributions establish language scent as a theoretical lens on multilingual information seeking and offer design implications for future search systems.

\begin{acks}
This work is supported in part by the National Science Foundation (NSF) under Award \#2230466.
We thank Linda Zeng for her help coding lab study information-seeking outcomes, all our participants for their time, and our reviewers for their valuable insights.
\end{acks}

\bibliographystyle{ACM-Reference-Format}
\bibliography{references.bib}

\appendix
\section{Likert Items in the Lab Study}
\label{app:likert}
\subsection{{Overall Impression }}
\begin{itemize}
    \item The multilingual aspect of this system did not provide any additional benefit compared to searching in a single language.
\end{itemize}
\subsection{Ease of Use}
\begin{itemize}
    \item I could search in multiple languages efficiently.
    \item I had to deal with too much information.
    \item Switching between languages during the information seeking process felt smooth.
    \item It was difficult to connect my intermediary search results, prior knowledge, and thoughts across different languages.
\end{itemize}

\subsection{{Forming Language Scent}}
\begin{itemize}
    \item The similarities and differences in information across languages were difficult to identify.
    \item It was easy to gauge the kinds of information available in each language.
    \item It was clear when searching in another language would be useful.
\end{itemize}
\subsection{{Reflecting on Language Scent}}
\begin{itemize}
    \item This system helped me evaluate the effectiveness of my multilingual information seeking strategies.
    \item This system supports reflection on my intuitions about the availability of information across languages.
\end{itemize}

{
\section{Full statistical tests in Lab Study}
\label{app:stats}

\subsection{Likert-scale Questionnaire}

\begin{table*}[tbp]
    \centering
    \caption{{\base{} and \sys{} Likert-scale statistics, based on Wilcoxon signed-rank test \cite{bridge1999IncreasingPhysiciansAwareness}.
    We applied the Benjamini-Hochberg correction to control for family-wise false discovery rate.
    We report the mean ($M$), adjusted and original $p$-values, and effect size ($r$).
    * and bolding indicates statistical significance.
    For italicized items, the smaller the better, for the rest, the greater the better.}
    }
        \begin{tabular}{p{7cm}ccccc}
        \toprule
        \textbf{Likert Item} & \textbf{$M_{\text{\base{}}}$} & \textbf{$M_{\text{\sys{}}}$} & \textbf{Adjusted $p$} & \textbf{Original $p$} & \textbf{$r$} \\
        \midrule
        \textit{The multilingual aspect of this system did not provide any additional benefit compared to searching in a single language.} & 2.75 & 1.94 & 0.303 & 0.061 & 0.583 \\ \midrule
        I could search in multiple languages efficiently. & 5.62 & 6.06 & 0.319 & 0.159 & 0.528 \\ \midrule
        \textit{I had to deal with too much information.} & 3.69 & 3.56 & 0.860 & 0.860 & 0.049 \\ \midrule
        Switching between languages during the information seeking process felt smooth. & 5.38 & 6.19 & 0.319 & 0.150 & 0.526 \\ \midrule
        \textit{It was difficult to connect my intermediary search results, prior knowledge, and thoughts across different languages.} & 3.75 & 2.88 & 0.319 & 0.140 & 0.432 \\ \midrule
        \textbf{*\textit{The similarities and differences in information across languages were difficult to identify.} }& 3.56 & 1.88 & 0.044 & 0.004 & 0.829 \\ \midrule
        It was easy to gauge the kinds of information available in each language. & 5.31 & 5.56 & 0.634 & 0.571 & 0.180 \\ \midrule
        It was clear when searching in another language would be useful. & 5.06 & 5.38 & 0.634 & 0.569 & 0.216 \\ \midrule
        This system helped me evaluate the effectiveness of my multilingual information seeking strategies. & 4.81 & 5.38 & 0.459 & 0.303 & 0.362 \\ \midrule
        This system supports reflection on my intuitions about the availability of information across languages. & 4.88 & 5.50 & 0.459 & 0.321 & 0.313 \\ \midrule
        \bottomrule
    \end{tabular}
    \label{tab:likert}
\end{table*}

\begin{figure*}[htbp]
    \centering
    \includegraphics[width = 1\linewidth]{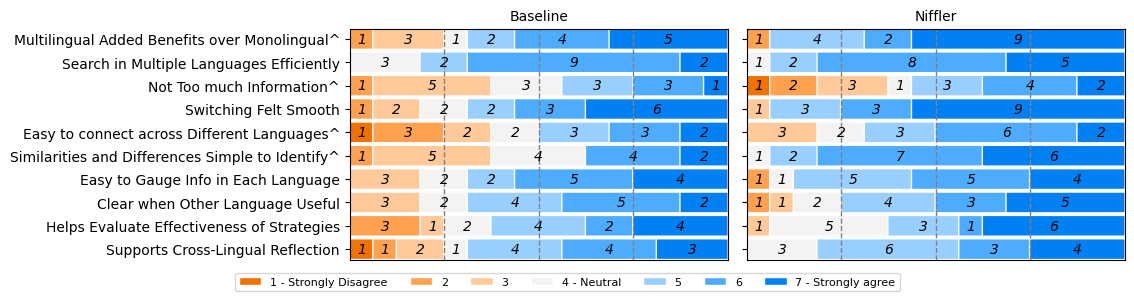}
    \caption{
    Distribution of Likert-scale Items from the Lab Study. Statements marked with \^{} are flipped for reporting. Original and full statements can be found in Appendix \ref{app:likert}}
    \label{fig:likert}
\end{figure*}

See Table \ref{tab:likert} and Figure \ref{fig:likert}.

\subsection{System Logs}
\label{sec:logStatNo}

See Table \ref{tab:log}. % and Figure \ref{fig:logStat}.

\begin{table*}[tbp]
\centering
\caption{{Statistical comparison between \sys{} and \base{} system logs.
We report the mean ($M$), $p$-value, and effect size ($r$) for each measure.
* and bolding indicates statistical significance.}
}
\begin{tabular}{lcccc}
\toprule
Measure & $M_{\text{\base{}}}$ & $M_{\text{\sys{}}}$ & $p$ & $r$ \\
\midrule
\textbf{*Number of queries} & 6.75 & 10.0 & 0.041 & 0.549 \\
Language switches & 1.44 & 1.94 & 0.401 & 0.221 \\
\textbf{*Language Engagement Span} & 0.657 & 0.403 & 0.019 & 0.601 \\
\textbf{*Language Balance} & 0.381 & 0.690 & 0.017 & 0.642 \\
Number of Sources Gathered & 8.13 & 9.44 & 0.404 & 0.207 \\
\textbf{*Number of topics explored} & 5.00 & 6.69 & 0.032 & 0.597 \\
\bottomrule
\end{tabular}
\label{tab:log}
\end{table*}
}

\section{Lab Study Method Details}
\subsection{Task Description}
\label{appendix:task}

\begin{enumerate}
    \item \textbf{Career.} You are teaching a career coaching course for university students who are about to graduate. In preparation, you want to gather a diverse set of career advice. Consider things like time management, goal setting, planning, choosing a career path, etc.
    \item \textbf{Food.} You are travelling to Switzerland for a month with a group of friends from different nationalities. You want to find as many foods and restaurants to try as possible, considering local specialties, the diverse tastes and dietary habits of your group, etc.
\end{enumerate}

\subsection{Measures}

\subsubsection{Information Seeking Process.}
\label{appendix:logMeasure}
We analyzed logs of user queries as a proxy for participants’ navigational behaviour during information seeking \cite{gwizdka2006WhatCanSearching, mathassan2005AssociatingSearchNavigation}.
Our focus was on high-level patterns of information seeking and language switching, rather than low-level details.
Specifically, we examined the following metrics: 
\begin{itemize}
    \item \textbf{Number of Queries}: The total number of queries conducted by a user during the session.
    \item \textbf{Language Switches}: The total number of times a user changed the query language during the session.
    \item \textbf{Language Engagement Span}: The average number of consecutive queries a participant conducted in the same language ($n_i$ for segment $i$), normalized by the total number of queries ($n$) in the session, i.e. $\frac{n_i}{n}$ for segment $i$.
    % For example, if the user first uses 5 queries in English (segment 1, $n_1 = 5$) followed by 1 query in Chinese (segment 2, $n_2 = 1$), the language engagement span is $\frac{\frac{5}{5 + 1} + \frac{1}{5 + 1}}{2} = 0.5$.
    The greater the span, the longer users stayed in the same language without switching, on average.
    \item \textbf{Language Balance}: Shannon entropy \cite{shannon1948MathematicalTheoryCommunication, lin1991DivergenceMeasuresBased} was used to quantify the balance of queries across languages. Higher values indicate a more even distribution of queries across languages, while lower values indicate skewness towards one language.
\end{itemize}

\subsubsection{Information Seeking Outcome}
\label{appendix:logCoding}

\begin{table*}[tbp]
\label{tab:topics}
\centering
\caption{The top 5 topics participants covered for each task, based on system logs.}
\begin{tabular}{ll}
\toprule
\textbf{Food} & \textbf{Career} \\
\midrule
food recommendation & career planning \\
swiss cuisine & career advice \\
restaurant recommendation & LinkedIn tips \\
Rösti\footnote{a swiss potato dish} & networking \\
international restaurants & curriculum vitae (CV) \\
\bottomrule
\end{tabular}
\label{tab:topics}
\end{table*}

For a proxy of the relevant information participants gathered, we used the sources participants clicked \cite{joachims2017AccuratelyInterpretingClickthrough}, or saved or took notes on \cite{zhengDiscipLinkUnfoldingInterdisciplinary2024a}.
In alignment with existing work, we used the number of topics as a measure of the diversity, or range of information covered \cite{zhang2020knowledgeUseChange, suh2023SensecapeEnablingMultilevelb}.
The topics were derived through systematic coding by two researchers, following the procedure in \cite{richardsPracticalGuideCollaborative2018}.
Each source was assigned the topic that most comprehensively describes its contents.
All coding was done blind to condition.
The coding rules were established by open coding and discussing 30\% of the data. 
As a pilot test, the two researchers independently coded 20\% of the data, achieving an initial percentage agreement \cite{halpinInterCoderAgreementQualitative2024a} of 69.6\%.
We judged percentage agreement based on whether the clustering aligned. For example, if both coders grouped the same three items together, we counted it as an agreement even if the cluster labels differed slightly (e.g., ``cheese dishes'' vs. ``cheese dish'').
After refining the coding rules and reaching agreement on the pilot test, the remaining 50\% of data were independently coded, this time reaching a percentage agreement of 86.5\%, above the conventional threshold of 80\% \cite{oconnorIntercoderReliabilityQualitative2020}.
The points of disagreement were discussed until resolved.
The top 5 topics for each task are presented in Table \ref{tab:topics} for reference.

\section{Technical Evaluation}
\label{app:techEval}
To further contextualize our study findings in terms of system capabilities, we conducted a technical evaluation of the generated comparative summaries.
From the original 268 queries created across participants during the tasks, duplicates were removed, resulting in 230 unique queries.
We then employed stratified sampling, randomly selecting 10 queries from each task (career, food) $\times$ language (English, Chinese) stratum, yielding a total sample of 40 queries along with the corresponding generated comparative summaries.
We recruited 2 experts who are fluent in both English and Chinese, and regularly seek information online.
They were asked to rate and comment on the components of the comparative summaries --- Cross-lingual Comparison, Summary of Sources in the original query language (L1), and Summary of Sources in the other language (L2), in relation to the query context.
Ratings were based on three dimensions: \textbf{accuracy} (binary; ``The summary is accurate.''), \textbf{answer relevance} (7-point Likert; ``Relevant information is provided.''), and \textbf{context relevance} (7-point Likert; ``No irrelevant information is provided.''), adopted from \cite{es2024RAGAsAutomatedEvaluation}.

On average, the comparison components had an accuracy of $87.5\%$, summaries of sources in the original query language (L1) had an accuracy of $100\%$ and summaries of sources in the other language (L2) had an accuracy of $97.5\%$.
Answer relevance ratings were $6.25$ (out of $7$) for the comparisons, $6.4$ for L1 summaries, and $6.15$ for L2 summaries.
Context relevance ratings were $6.18$ for the comparisons, $6.33$ for L1 summaries, and $6.03$ for L2 summaries.
Overall, the comparative summaries appeared reasonably accurate and relevant.
Issues mainly occurred due to (1) meaning lost in translation, particularly in query interpretation, and (2) the information provided or the point of comparison being overly generic or overly specific.

\end{document}